\documentclass[aps,prl,preprint]{revtex4}
\usepackage{graphicx}       
\usepackage{dcolumn}        
\usepackage{bm}             

\begin{document}


\title
    {
        Ultrafast imaging of photoelectron packets generated from graphite surface
    }
\author{Ramani K. Raman}
\affiliation{Physics and Astronomy Department,
             Michigan State University,
             East Lansing, Michigan 48824-2320 }

\author{Zhensheng Tao}
\affiliation{Physics and Astronomy Department,
             Michigan State University,
             East Lansing, Michigan 48824-2320 }

\author{Tzong-Ru Han}
\affiliation{Physics and Astronomy Department,
             Michigan State University,
             East Lansing, Michigan 48824-2320 }

\author{Chong-Yu Ruan}
\email[]{Email: ruan@pa.msu.edu}
\affiliation{Physics and Astronomy Department,
             Michigan State University,
             East Lansing, Michigan 48824-2320 }


\begin{abstract}

We present an electron projection imaging method to study the ultrafast evolution of photoelectron density distribution and transient fields near the surface. The dynamical profile of the photoelectrons from graphite reveals an origin of a thermionic emission, followed by an adiabatic process leading to electron acceleration and cooling before a freely expanding cloud is established. The hot electron emission is found to couple with a surface charge dipole layer formation, with a sheet density several orders of magnitude higher than that of the vacuum emitted cloud.

\end{abstract}



\maketitle



Understanding the mechanisms of vacuum space charge (VSC) \cite{ Passlack2006, Zhou2005, GiltonJAP90} emission and surface charge formation is central to the development of pulsed laser driven electron technologies such as time-resolved photoemission \cite{Passlack2006, Zhou2005, Kubo05}, scanning probe microscopy \cite{MeyerzuHeringdorf07, Gerstner2000}, electron diffraction \cite{Srinivasan2003, MillerFEDRvw06, Park2005, Hanisch2008, ZewailUEDRvw2006, RuanMM09}, and microscopy \cite{Lobastov2005, KimDTEM08}. While the vacuum space charge has been studied both theoretically \cite{Hilbert09, ReedJAP06, SiwickJAP02, Qian2002} and experimentally \cite{Hebeisen08, Passlack2006, Zhou2005,Srinivasan2003}, the role of subsurface electron dynamics within the materials and the nature of the early development of VSC remain topics of high interest. Recently, the transient vacuum electric field established by the generation of VSC has been investigated, based on the influence of the field on a pulsed electron beam \cite {ParkAPL09, HebeisenPRB08}. Here, a method to directly image the spatiotemporal evolution of the photo-emitted electron bunch generated over a femtosecond laser excited surface is presented, based on an electron projection geometry. The method possesses sufficient sensitivity to image electron bunches as small as $10^{10}$ e/cm$^3$ and permits quantitative measurement of the instantaneous electron bunch density distribution and its translational and expansion velocities in the picosecond and micrometer regime. Contrary to the \emph{space-charge} heating, we observe an adiabatic cooling during the initial expansion of the thermally emitted electrons, causing a nearly 80$\%$ drop of the internal temperature while accelerating a high CoM velocity. In conjunction with the diffractive voltammetry \cite{RuanMM09, MurdickPRB08}, this offers a way to study simultaneously both subsurface charge dynamics as well as vacuum emitted space charge effects self-consistently.

\begin{figure}
\includegraphics[width=0.75\columnwidth]{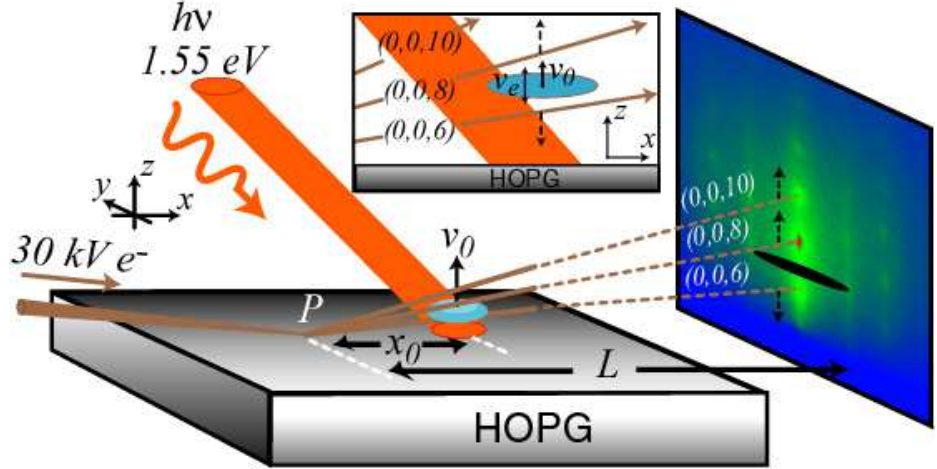}
\caption{ (Color online) Schematics of the electron point-projection imaging
					technique. Typical values are $x_0\approx$ 1 - 4 mm, $L\approx$
					150 mm. The dashed arrows illustrate the effect of the photo-emitted
					electron bunch on the Bragg beams.	}	
				\label{Fig1}
\end{figure}

The ultrafast electron imaging is conducted in a pump-probe experimental arrangement, in which the femtosecond laser (800 nm, 45 fs) is used to generate the photoemission from graphite (HOPG, ZYA grade, SPI Supplies), which is subsequently probed by the delayed surface scattered electrons. The incident laser beam is displaced from the source of electron scattering ($P$) by a distance $x_{0}$ (typically a few mm) to establish a projection imaging geometry (Fig. \ref{Fig1}). The presence of VSC diminishes the forward-scattered electrons generated at $P$, thus casting a shadow on the screen, which is at a distance $L=$~150 mm away from $P$. The magnification from this projection imaging is $M \approx L/x_{0} \approx 100$. By adjusting the arrival time of the probe electrons relative to that of the laser, shadow images of the evolving electron bunch can be obtained, as shown in Fig. 2. The cross-sectional line scans obtained from these shadow images reveal an accumulation peak near the shadow edge (caused by surface scattering), which decays sharply into the vacuum, followed by a depletion profile characteristic of a gaussian evolving in space over time. To extract the spatiotemporal dynamics of the electron bunch, we fit the line scans to the following analytical form, which takes into account the effect of the projection geometry:

\begin{figure}
	\includegraphics[width=1.0\columnwidth]{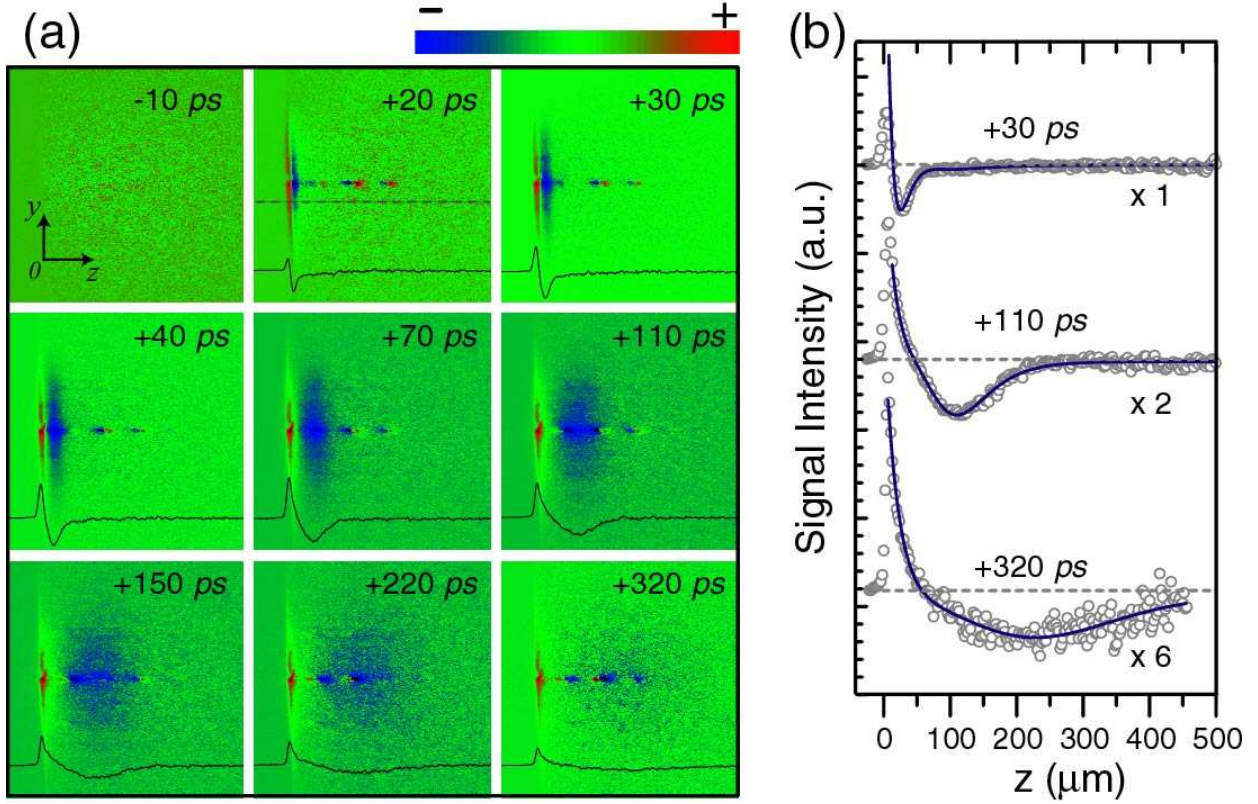}
	\caption{ (Color online)
						(a)	Snapshot shadow images of photoemitted electron bunch
								obtained with $x_{0}=$ 4.3 mm. The cross-sectional
								line profiles (black curves) are extracted along the dashed
								line in the second panel.
						(b) Fitting curves (lines) of cross-sectional line-profile
								data (circles) using a gaussian model considering the projection geometry.
	\label{Fig2}}
\end{figure}

\begin{equation}
	 F(d) = 	A\:exp(-d/\Delta_0) +
					B\frac{\Sigma_0}{\Delta_z}
							exp \left( - \frac{	\frac{(dx_0-Lz_0)^2}{2d^2\Delta_x^2 + L^2\Delta_z^2}
																}{\sqrt{\frac{1}{\Delta_x^2} + \frac{d^2}{L^2\Delta_z^2}}}
					 		 			\right)
	\label{Eq:1}
\end{equation}

\noindent where $d$ is the distance on screen measured from the shadow-edge. The charge distribution $\rho(z,t)$ is modeled as gaussians in $x$, $y$ and $z$ directions with a sheet electron density $\Sigma_0$ in the $xy$-plane, 1/$e$ half-widths $\Delta_x$, $\Delta_y$, $\Delta_z$ and a CoM position ($x_0$, $y_0$, $z_0$). $\Delta_x \approx$ 500 $\mu m$ is determined by examining the transverse size of the electron bunch, which remains nearly unchanged and corresponds well with the width of laser illumination on the surface. The evolution of the bunch in the $z$-direction is shown in Figs. \ref{Fig3} (a), (b) and summarized in Table I for fluences $F=$7, 23 and 56 mJ/cm$^2$. The results indicate the electron cloud CoM to follow a projectile-like trajectory (caused by the attractive force from image charges, Fig. \ref{Fig3}(a)), with a translational CoM velocity $v_0=dz_0/dt$, while undergoing free-expansion as indicated by the nearly constant expansion speed $v_e=d\Delta_z/dt$ in Fig. \ref{Fig3}(b).

From the gaussian shape of the velocity distribution and the lack of a power-law enhancement of $\Sigma_0$ over increasing $F$, we believe a thermionic emission scenario is best to describe the dynamical parameters observed here.  For a steady-state thermal emission, the initial translational speed $v_i$ and the expansion speed $v_e$ of the bunch is related to the electronic temperature ($T_e$) at the surface via the relations: $v_i= \sqrt{k_BT_e/2\pi m_e}$ , and $v_e=\sqrt{k_BT_e/m_e}$. The observation of a $v_0$ significantly higher than $v_e$ in Table \ref{table:T1} suggests an electron acceleration and cooling during the initial adiabatic expansion of the cloud. This scenario is supported by the observation of an apparent shift of the \emph{zero-of-time} extracted from the linear extrapolation of cloud expansion to intercept with the time axis (Fig. 3(b)), indicating a rapid decrease of $v_e$ within 10 $\mu$m (resolution-limited) of the expansion. This cooling process converts part of the electronic enthalpy ($\Delta h=5/2k_B\Delta T$) to the kinetic energy of the bunch, and under such an adiabatic model we can deduce the initial electron temperature $(T_i)$ using $1/2m_e(v_o^2-v_i^2)=\Delta h$. The $T_i$ deduced from our experiment shows a saturation at high fluence, and the highest electronic temperature obtained is just above the laser pulse energy 1.55 eV (18,000 K), further confirming a thermally limited emission. During the same period, by assuming an initial cloud size of 30 nm (comparable to the laser penetration depth) and $\Sigma_0$ = 2 $\times$ $10^{8}$ e/cm$^{2}$, we estimate the possible heating effect arising from space-charge driven acceleration of the electrons at the expense of Coulomb self-energy to be at most 0.2 eV. Thus Coulomb explosion cannot provide the observed high initial CoM velocity.

\begin{figure}
	\includegraphics[width=1.0\columnwidth]{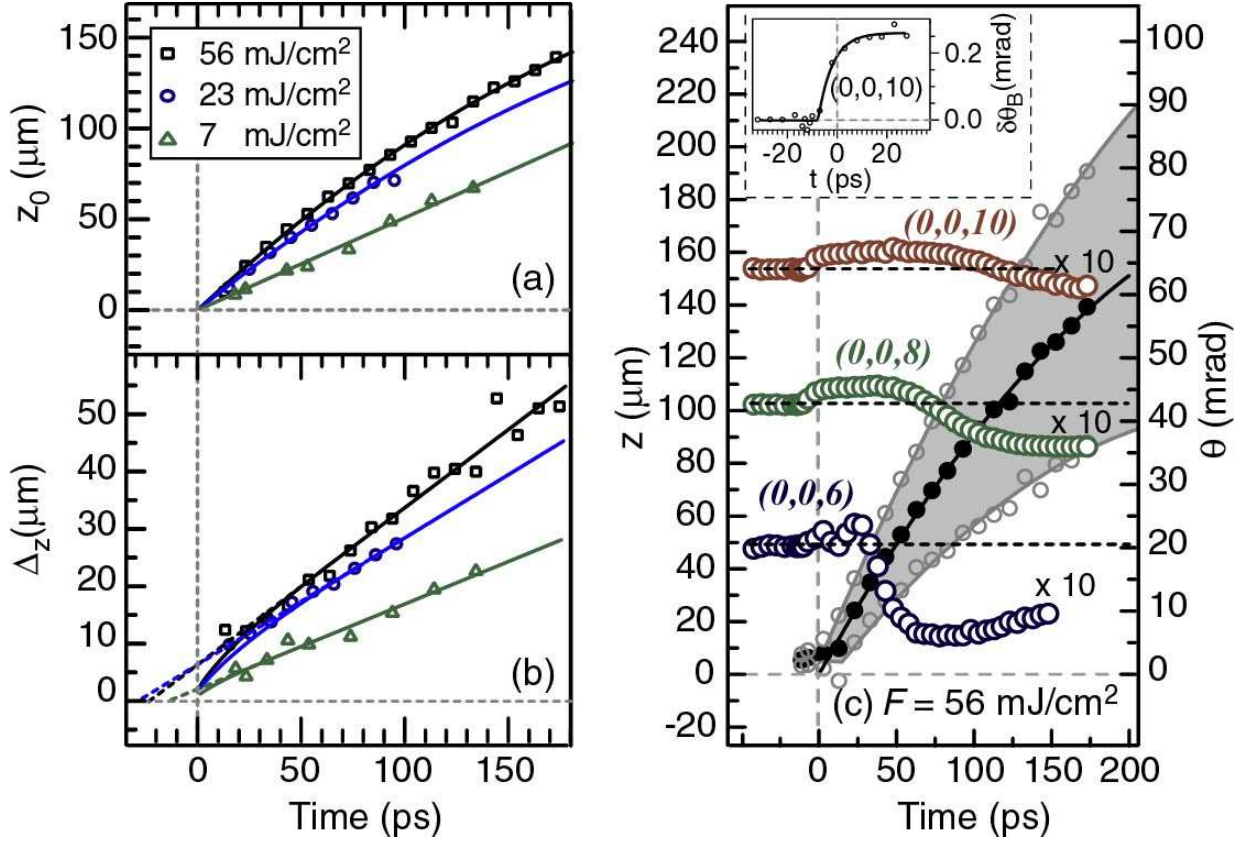}
	\caption{ (Color online)	
			Photoemission from graphite surface: Dynamics along the
			$z$-direction, when $x_{0}=$ 2.4 mm.
			(a) Time-evolution of the electron bunch's center-of-mass (CoM) and
			(b) 1/$e$-width.
			(c) Spatiotemporal evolution of the Bragg peaks and the electron bunch
					at $F=$ 56 mJ/cm$^2$, showing the bunch CoM (black dots) and its
					spread (1/$e$-width) relative to the CoM (gray circles).
			}
	\label{Fig3}
\end{figure}

In addition to the imaging, we also analyze the trajectory of the diffracted electron beams (0,0,6), (0,0,8), (0,0,10) in the central streak region of the diffraction pattern. The instantaneous vacuum electric field $E(z,t)$ established by the two opposing fields associated with the VSC $\rho(z,t)$ and its mirror-charges on the surface can be described by :
\begin{equation}
	E(z,t) = \frac{e}{2\epsilon_0} 	
					\left( a(z)\Sigma_0 -
									2 \int\limits_{z}^\infty \rho(z',t)dz'
					\right)
	\label{Eq:2}
\end{equation}
\noindent where $a(z)\approx z/\sqrt{z^2+\Delta_x^2}$ is the proximity effect factor\cite {ParkAPL09}, describing the reduction in local electric field from mirror charges with a finite-sized slab, which is [1-$a(z)$]e$\Sigma_0/(2\epsilon_0)]$. The Bragg beams located at a height $z_c$ above the surface interact with VSC and exhibit an inversion, as shown in Fig. \ref{Fig3}(c) when the CoM of the electron bunch reaches a height $z_0=z_c-\gamma \Delta_z$, such that the local electric field at $z_c$, $E(z_c)=0$. Carrying out the integration in Eqn. \ref{Eq:2}, we obtain the condition for Bragg beam inversion as $\gamma=1.6$, 1.3, 1.1 for $z_0=$ 50 $\mu m$, 100 $\mu m$ and 150 $\mu m$ corresponding to the (0,0,6), (0, 0, 8) and (0, 0, 10) diffracted beams respectively. Inspecting the trajectories $\Delta_z(t)$ and $z_0(t)$ in Fig. 3(c) at the point of Bragg beam inversion, we measure $\gamma=$ 1.0 $\pm$ 0.4, 1.2 $\pm$ 0.2 and 1.2 $\pm$ 0.2, which is in good agreement with the predicted values above. The deviation in the case of (0,0,6) beam is believed to be caused by the recoil effect, which plays a more significant role here. Given this confirmation, we can reliably obtain the sheet electron density $\Sigma_0$ for the three fluences studied here (Table I), and compare them with the initial electron temperature $T_i$. $\Sigma_0$ scales linearly with $T_i$ to a satisfactory degree, agreeing with the thermionic emission model proposed by Downer and coworkers \cite{Riffe1993}. The electron bunch dynamics measured here reveals that the majority of the electrons will return to the surface within 1 ns, see $t_{return}$ in Table I, with only a very small portion ($<10^{-4}$) of the cloud remaining in vacuum beyond 3 ns.

\begin{table}
	\caption{
						Results of fitting experimental data to the
						projected gaussian model.
					}
	\small
			\begin{tabular}{c c c c c c c}
			\hline\hline
			
				$	F		$					& 	
				$	v_0	$					&
				$	v_e	$					&	
				$	T_i	$					&
				$	T_e	$					& 		
				$\Sigma_0$			&
				$t_{return}	$		\\
				
												&
				$\times 10^6$ 	&
				$\times 10^6$ 	&
				$\times 10^3$		&
				$\times 10^3$		&
				$\times 10^8$		&
												\\
				
				(mJ/cm$^2$)			&
				m/s 						&
				m/s				 			&
				K								&
				K								&
				e/cm$^2$				&
				ps							\\
				
			\hline
			
				56			& 		1.06 		&			0.27 		&	
				19			& 		5.0			& 		1.93		&
				350 		\\
				
				23			&			0.91		&			0.22		&
				14			&			3.3			&			1.61		&	
				400 		\\
				
				7				&			0.51		&			0.15 		&
				4.7			&			1.5			&			0.39		&
				820 \\

			\hline\hline
		\end{tabular}
	\label{table:T1}
\end{table}

\begin{figure}
	\includegraphics[width=1.0\columnwidth]{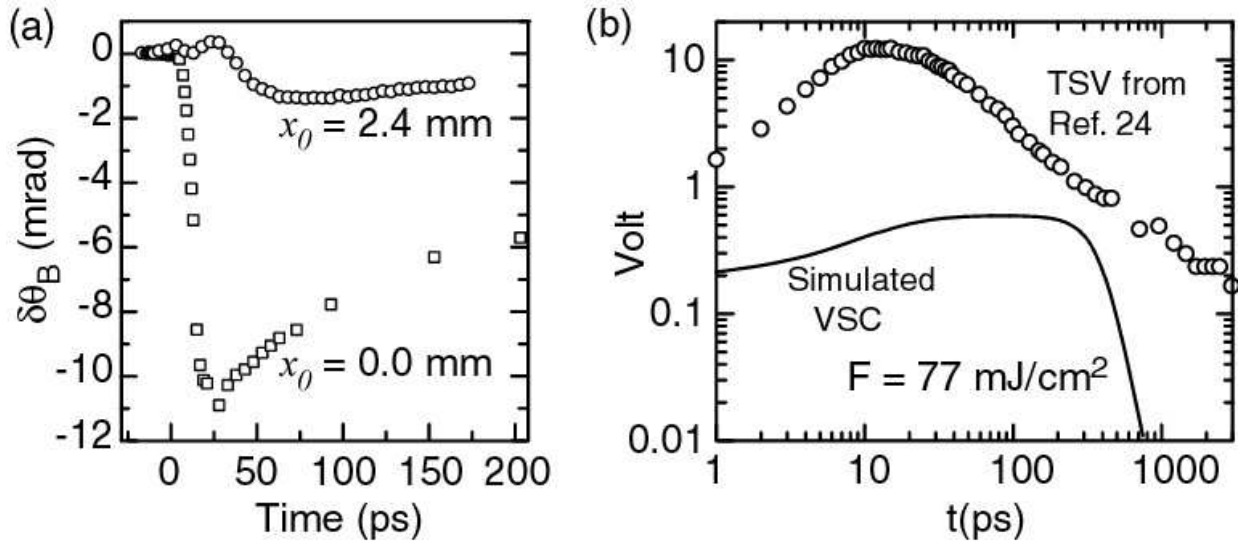}
	\caption{ Comparision of TSV and VSC effects in UEC investigations of HOPG.
						(a) Shift of the (0,0,6) Bragg peak in case of perfect
								pump-probe alignment ($x_{0}=$~0) and projection imaging geometry
								with ($x_{0}=$~2.4 mm).
						(b) Comparison of the TSV measured by the probing electron
								beam submerged beneath the HOPG surface in the UEC study (Ref. \cite{RamanPRL08}) along with the estimated effect of VSC in that study, based on the VSC dynamics extracted here.
	 }
	\label{Fig4}
\end{figure}

We also observe a dramatic increase in the electron refraction ($\delta \theta_B$) \cite{MurdickPRB08} simply shifting the laser to $P$ without changing the diffraction geometry, as shown in Fig. \ref{Fig4}(a). Simulation of the VSC induced refraction effect in the case of perfect pump-probe alignment (Fig. \ref{Fig4}(b)) clearly indicates that VSC alone is not sufficient to explain the observed large and rapid 10 V rise in the transient surface voltage of fs-laser excited graphite surface \cite{RamanPRL08}. This consequently mandates the presence of a surface dipole field that is invisible in vacuum to account for the full refraction effect. Assuming a probe-electron depth of $\approx$1 nm, we deduce a ps retention of surface charge density in graphite on the order of $10^{14}$ e/cm$^2$, which is comparable to the typical surface/interface state density. The thermionic emission on a sub- to a few picosecond timescale is likely mediated by the image/interlayer states of graphite, which has a strong 3D character\cite{SilkinPRB2009}. These states above the vacuum level can efficiently transport electrons to vacuum and create the dipole field near the surface. Following the electron cooling, the coherent interlayer transport normal to the basal planes of graphite is essentially turn off, causing the slow decay of the surface dipole field, which plays a major role in influencing the surface structure dynamics\cite{RamanPRL08, Kanasaki2009, CarboneScience2009}. In conjunction with transient surface voltammetry, this vacuum electron imaging method can be applied to investigate a variety of hot electron processes involving intense laser interaction with solid surfaces.

We acknowledge Ryan Murdick and Pampa Devi for for helpful discussions. This work was supported by Department of Energy under grant
DE-FG02-06ER46309.

 \bibliographystyle{apsrev}

\end{document}